\documentclass[review]{elsarticle}
\usepackage{lineno,hyperref}
\usepackage{listings}
\usepackage{textcomp}
\usepackage{adjustbox}
\usepackage{multirow}
\usepackage{dingbat}
\usepackage{amsmath}
\usepackage{array}
\usepackage{booktabs}

\modulolinenumbers[1]

\journal{Journal of Systems and Software}

\begin{document}

\begin{frontmatter}

\title{ACCORDANT: A Domain Specific Model and DevOps Approach for Big Data Analytics Architectures\tnoteref{ack}}
\tnotetext[ack]{This research was carried out by the Center of Excellence and Appropriation in Big Data and Data Analytics (CAOBA), supported by the Ministry of Information Technologies and Telecommunications of the Republic of Colombia (MinTIC) through the Colombian Administrative Department of Science, Technology and Innovation (COLCIENCIAS) within contract No. FP44842-anexo46-2015. We would like to make an special acknowledgement to the Taxation Authority of Bogota at the Bogota's Department of Finance (SHD), to the office of telecommunications of the National Planning Department (DNP), and to the CAOBA-Uniandes team, both faculty and students.}


\author[mainaddress]{Camilo Castellanos\corref{mycorrespondingauthor}}
\cortext[mycorrespondingauthor]{Corresponding author}
\ead{cc.castellanos87@uniandes.edu.co}

\author[secondaryaddress]{Carlos A. Varela}
\ead{cvarela@cs.rpi.edu}

\author[mainaddress]{Dario Correal}
\ead{dcorreal@uniandes.edu.co}

\address[mainaddress]{System Engineering and Computing Department, University of Los Andes, Bogota, Colombia}
\address[secondaryaddress]{Department of Computer Science Rensselaer Polytechnic Institute, Troy, NY, USA}

\begin{abstract}
Big data analytics (BDA) applications use machine learning algorithms to extract valuable insights from large, fast, and heterogeneous data sources. 
New software engineering challenges for BDA applications include ensuring performance levels of data-driven algorithms even in the presence of large data volume, velocity, and variety (3Vs).
BDA software complexity frequently leads to delayed deployments, longer development cycles and challenging performance assessment.
This paper proposes a Domain-Specific Model (DSM), and DevOps practices to design, deploy, and monitor performance metrics in BDA applications. Our proposal includes a design process, and a framework to define architectural inputs, software components, and deployment strategies through integrated high-level abstractions to enable QS monitoring. We evaluate our approach with four use cases from different domains to demonstrate a high level of generalization.
Our results show a shorter deployment and monitoring times, and a higher gain factor per iteration compared to similar approaches.
\end{abstract}

\begin{keyword}
Software architecture \sep Big data analytics deployment \sep DevOps \sep domain specific model \sep quality scenarios \sep performance monitoring 
\end{keyword}

\end{frontmatter}


\section{Introduction}

Big data analytics (BDA) applications use Machine Learning (ML) algorithms to extract valuable insights from large, fast and heterogeneous data. These BDA applications require complex software design, development, and deployment to deal with big data characteristics: volume, variety, and velocity (\textit{3Vs}), to maintain expected performance levels. Specifically, BDA processing takes advantage of cutting-edge technologies and infrastructures that enable distributed stream computing. But the complexity involved in BDA application development frequently leads to delayed deployments \cite{Chen2016} and hinders performance monitoring (e.g. throughput or latency) \cite{Ranjan2014}. Regarding the 3Vs, a BDA solution can be constrained to different performance metrics. For instance, real-time stream analytics applications require low latency and flexible scalability based on data volume fluctuation. On the other hand, heavy workloads, which imply batch processing over big data, demand high scalability and fault tolerance to achieve a particular deadline. One of the key goals of software architecture is the design of the system's structures and their relationships to achieve expected quality properties.

The development of BDA solutions involves three knowledge domains: business, analytics, and technology. In the business domain, business experts have to define business goals and quality scenarios (QS) to drive analytics projects. 
In the analytics domain, these business goals are translated into specific analytics tasks by data scientists. Finally, in the technology domain, software architects make decisions in terms of tactics, patterns, and deployment considerations keeping in mind quality attributes. Stakeholders from different domains face heterogeneous concerns and different abstraction levels. Due to the lack of techniques, and tools to enable articulation and integration of such domains, BDA solutions development presents a high cost and error-prone transition between development and production environments \cite{Chen2016,Wegener2011}. Though there is a growing interest of companies in big data adoption, real deployments are still scarce (``Deployment Gap'' phenomenon) \cite{Chen2017}.

In the same vein, previous surveys \cite{Rexer2013,Rexer2016,Castellanos2019} have reported low deployment frequency and delayed deployment procedures caused by analytics model translation, lack of tools' interoperability and stakeholders' communication. These pitfalls could be the result of the traditional approach of BDA development where the data scientist produces the models as source code implemented using machine learning-oriented tools which are focused on analytics perspectives within a controlled environment (data lab). On the other hand, software architects have to translate these models into software products which usually implies rewriting code to obtain productive software components deployed on specific IT infrastructures. 

This paper proposes ACCORDANT (An exeCutable arChitecture mOdel foR big Data ANalyTics), a DevOps and Domain-Specific Model (DSM) approach to develop, deploy, and monitor BDA solutions bridging the gap between analytics and IT domains. ACCORDANT allows to design BDA applications using QS, functional, and deployment views. A QS specifies a quality attribute requirement for a software artifact to support design and quality assessment. Functional view defines the architectural elements that deliver the application's functionality. Deployment view describes how software is assigned to hardware-processing and communication elements. Our deployment strategy incorporates containerization since it offers consistent modularity to facilitate portability, continuous integration, and delivery.

ACCORDANT is validated with four use cases from different domains by designing functional and deployment models, and assessing performance QS. This validation aims to reduce the time of design, deployment, and QS monitoring of BDA solutions. 
These use cases range from public transportation and avionics safety to weather forecasting and they include distributed batch, micro-batch, and stream processing. Our results indicate improvements in design and (re)deployment times to achieve the expected performance metrics. In summary, the contributions of this paper are as follows:

\begin{itemize}
    \item A DSM framework to formalize and accelerate iteratively the development and deployment of BDA solutions by specifying architectural functional, and deployment views aligned to QS.
    \item Three integrated domain-specific languages (DSLs) to specify architectural inputs, component-connector models, and deployments, thus accelerating BDA deployment cycle.
    \item A containerization approach to promote automation delivery and performance metrics monitoring for BDA applications aligned to QS.
    \item The evaluation of this proposal applied to four use cases from diverse domains, and using different deployment strategies and QS.
\end{itemize}

The rest of this paper is organized as follows. In Section \ref{sec:background}, we describe the background on DSM, big data analytics, and DevOps. Section \ref{sec:relatedwork} reviews related work. Section \ref{sec:proposal} presents our methodology and proposal overview. Section \ref{sec:evaluation} presents the use cases for experimentation. Section \ref{sec:experimentation} illustrates the steps followed to validate this proposal. Section \ref{sec:results} presents and discusses the obtained results. Finally, Section \ref{sec:conclusions} summarizes the conclusions and future work.

\section{Background}
\label{sec:background}

This section describes the core concepts in which this proposal is supported: domain-specific modeling, software architecture, big data analytics, and Dev\-Ops. 

\subsection{Domain-Specific Modeling (DSM) and Software Architecture}

Domain-Specific Modeling enables the software to be modular and resilient to changes through the separation of concerns (SoC) principle by specifying technology-agnostic concepts, relationships, and constraints within the domain. An important advantage of DSM is the close mapping problem and solution domains to provide code generation. Moreover, DSM can speed up and optimize the code generated for the specific platform improving productivity. In order to enable code generation, the domain model requires to be narrow, and it is constrained by a language specification, the \textit{metamodel}. Furthermore, due to the narrow metamodel's scope, the models can be read, checked, validated, and interpreted to generate specific implementations.
Regarding representations, DSM can be expressed in graphical, textual, or mixed notation according to the domain context. It is possible to embed multiple views or aspects (for example, analytics, software components, and deployment) using different representations that share elements or mappings. 

An architecture description language enables architects to express high-level system structure by describing its coarse-grained components and connections among them. 
These descriptions are contained in \textit{architectural views} to address different concerns, and these views are built based on collection of patterns, templates, and conventions called \textit{Viewpoints} \cite{Rozanski2005}. The architectural design is driven by quality scenarios and primary functional requirements through a systematic design method, such as the Attribute-Driven Design method (ADD \cite{cervantes2016}). ADD starts identifying inputs: QS, functional requirements, and constraints. In each ADD iteration, a design goal is defined from these inputs, and the selection of architectural structures, tactics, patterns, and their application described across views, aims at achieving such goal.
A pattern is a standard, known and reusable solution to a common problem in software architecture.
Tactics are design primitives to achieve a response for particular quality attributes. Previous studies have collected both patterns \cite{erl2016,Marz2015} and tactics \cite{Gorton2014,ullah2019} to be applied in the BDA domain.

\subsection{Big Data Analytics}

In BDA context, data processing models aim at specific application requirements: \textit{batch} to process large stored datasets all at once with high performance, and \textit{stream} processing for an unbounded data flow in (near) real-time. Due to the complexity of deploying and operating BDA solutions integrating a myriad of technologies, complex analytics models and distributed infrastructure, some research has been done to tackle such complexity by raising the level of abstraction \cite{Gribaudo2017, Guerriero2016,Huang2015}.

Due to the wide range of BDA technologies, portability plays a key role to deploy, operate, and evolve BDA applications, and this is where portable standards appear such as Predictive Model Markup Language (PMML)\footnote{http://dmg.org/pmml/v4-3/GeneralStructure.html} or Portable Format for Analytics (PFA)\footnote{http://dmg.org/pfa/}. PMML is the de facto standard proposed by the Data Mining Group that enables portability of analytics models through neutral-technology XML format. PMML allows specifying a set of machine learning models and data transformations along with their metadata.

\subsection{DevOps and IaC}
According to Bass. et. al \cite{bass2015}, DevOps is a set of practices aims to reduce the time from software development to production environment, ensuring high quality. DevOps includes activities as deploy, operate and monitor applications, with the goals of improve deployment frequency and speed up the time to market what is aligned to our proposal's objectives. Infrastructure as Code (IaC) arises from the necessity to handle the infrastructure setup, evolution, and monitoring in an automated and replicable way through executable specifications.
IaC promotes the reduction of cost, time and risk of IT infrastructure provision by offering languages and tools which allow to specify concrete environments (bare-metal servers, virtual machines, operative systems, middleware and configuration resources) and allocate them automatically. In this context, technologies such as Kubernetes\footnote{https://kubernetes.io/}, an open
source for automating deployment, scaling, and management of container clusters which offers to decouple application containers from the infrastructure details.

\section{Related Work}
\label{sec:relatedwork}

Several works have proposed frameworks to build and deploy BDA applications. We review and compare some of the most relevant works, that comprise building blocks to construct and deploy BDA pipelines. Indeed, some works have tackled DSM to describe functional and deployment viewpoints involving DevOps practices. We summarize and compare the related work reviewed in Table \ref{tab:rw}, addressing the identified problem and our vision of using separation of concerns (\textit{SoC}), domain-specific modeling and DevOps to deal with the deployment gap.

Table \ref{tab:rw} details in each column some features we identify in the related work as follows. \textit{SoC} is a key design principle for us, since the knowledge domains involved in BDA (business, analytics, and IT) have to be tackled from different perspectives (i.e. viewpoints). In terms of analytics domain, cross-industry (\textit{CI}), and technology-neutral models (\textit{TNM}) promote applicability, and BDA portability respectively. Regarding software architecture concepts, QS specification (\textit{QSS}), functional (\textit{FV}) and deployment (\textit{DV}) views allow us to describe orthogonal concerns such as quality scenarios, components-and-connector, and deployment models. Architectural tactics (\textit{AT}) are design decisions that influence the control of a QS response. A target-technology assignment (\textit{TTA}) complements DSM approaches by supporting a predefined technologies set (\textit{P}) or extensible code generators (\textit{C}). Finally, considering the DevOps practices, deployment specification column (\textit{DS}) defines if only a number of instances (\textit{I}) per component or a whole deployment diagram (\textit{D}) can be described. Additional practices that facilitate the deployment and operation processes are considered: continuous deployment (\textit{CD}), QS monitoring (\textit{QSM}), and self-adaptation (\textit{SA}). 

\begin{table}[]
\caption{Related Work}
\label{tab:rw}
\begin{adjustbox}{max width=\textwidth}
\begin{tabular}{|p{3.3cm}|c|cc|ccccc|cccc|}
\cline{1-13}
\multirow{2}{*}{Work} & \multicolumn{1}{l|}{\multirow{2}{*}{SoC}} & \multicolumn{2}{c|}{Business (Analytics)} & \multicolumn{5}{c|}{Software Architecture} & \multicolumn{4}{c|}{DevOps}  \\
\cline{3-13}
 & \multicolumn{1}{l|}{}  & CI & \multicolumn{1}{c|}{TNM} & QSS & FV & DV & AT & \multicolumn{1}{l|}{TTA} & DS & CD & QSM & \multicolumn{1}{l|}{SA} \\
\hline
Lechevalier et al. \cite{Lechevalier2015} & & \checkmark & \checkmark & & \checkmark & & & & & & &  \\
Gribaudo et al. \cite{Gribaudo2017}, Huang et al. \cite{Huang2015} & & \checkmark & & \checkmark &  & & & & D & & \checkmark &  \\
CloverDX \cite{CloverDX} & & \checkmark & & & \checkmark & & & & I & \checkmark & \checkmark &  \\
OptiML \cite{Sujeeth2011} & & \checkmark & \checkmark & & \checkmark & & & C & & \checkmark & & \\
Qualimaster \cite{Alrifai2014} & \checkmark & & & \checkmark & \checkmark & \checkmark & \checkmark & & & \checkmark & \checkmark & \checkmark \\
FastScore \cite{OpenDataGroup} & \checkmark & \checkmark & \checkmark & & \checkmark & & & C & I & \checkmark & \checkmark & \\
SpringXD \cite{Anandan2015} & \checkmark & \checkmark & \checkmark & & \checkmark & & & P & I & \checkmark & \checkmark & \checkmark \\
DICE \cite{Guerriero2016}\cite{Artac2018}\cite{perez2019} & \checkmark & \checkmark & & \checkmark & \checkmark & \checkmark & & C & D & \checkmark & \checkmark & \checkmark\\
\hline
\textbf{ACCORDANT} & \checkmark & \checkmark & \checkmark & \checkmark & \checkmark & \checkmark & \checkmark & C & D & \checkmark & \checkmark & \\
\hline
\end{tabular}
\end{adjustbox}
\end{table}

Some works have presented DSM to model analytics functions, however, they do not tackle architecture concepts and deployment considerations because they are only focused on functional definitions. Lechevalier et al. \cite{Lechevalier2015} introduce a DSM framework for predictive analytics of manufacturing data using artificial neural networks to generate analytics models. Sujeeth et al. present in \cite{Sujeeth2011} OptiML, a DSL for machine learning which describes analytics functions using a statistical model which cover a subset of ML algorithms, this analytics functions are analyzed and optimized before the code generation. CloverDX \cite{CloverDX} is a commercial tool to design data transformations and analytics workflows in a visual way integrating external APIs, and including parallel processing in multiple nodes. CloverDX's functional view includes readers, processors, and writers for a predefined set of technologies, but deployment view is not available and distributed processing must be defined with specific parallel nodes in the functional view, which prevents to use the same functional definition in different deployment strategies. Finally, technology-neutral models, performance scenario specifications, and architectural tactics are not supported. 

In contrast, we found another group of studies interested in infrastructure concerns of BDA applications leaving aside their functional components. Gribaudo et al. \cite{Gribaudo2017} propose a modeling framework based on graph-based language to evaluate the system's performance of running applications which follow the lambda architecture pattern. This modeling framework allows users to define stream, batch, storage, and computation nodes along with performance indices to be simulated and evaluated, but neither functional BDA application nor real infrastructure provision are provided as a result. Huang et al. \cite{Huang2015} introduce a model to design, deploy, and configure Hadoop clusters through architecture metamodel and rules, which describe BDA infrastructure and deploy automation. Their work is focused on design, deployment, and evaluation of BDA technology infrastructures. However, it leaves out functional analytics models to get an integrated BDA solution.

QualiMaster \cite{Alrifai2014} focuses on the processing of online data streams for real-time applications such as the risk analysis of financial markets regarding metrics of time behavior and resource utilization. The aim of QualiMaster is to maximize the throughput of a given processing pipeline. Similarly, our proposal generates software for BDA applications, but taking as input the analytics specification of a predictive model, and the performance metrics to be achieved. Unlike Qualimaster, our proposal is technology-neutral and cross-industry which enables a more widespread application.

Fastscore \cite{OpenDataGroup} is a commercial framework to design and deploy analytics models. Analytics components are conventionally developed using a determined programming language or using a PMML file, and once imported to the platform, they can be connected to data inputs and outputs. Quality scenarios cannot be specified, but performance metrics can be visualized. Deployment is realized through \textit{engines} (containers) where models are executed, and the deployment design is limited to engine replication factor to increase the concurrency of analytics models.

SpringXD \cite{Anandan2015} is a unified, distributed, and extensible system for data ingestion, analytics, processing, and export to simplify BDA development and deployment. In SpringXD, modules are data processing units of one of three types: \textit{source}, \textit{processor}, or \textit{sink}, and they can be connected using messaging abstractions called \textit{message bus} to build BDA pipelines. Modules run over a cluster of containers which can be replicated to a fixed number and monitored to observe performance behavior, although these metrics are not application-oriented, but infrastructure-oriented (e.g. CPU and memory use). Similar to our approach, analytics processor can be defined through PMML models, but target technologies are limited to a set of predefined options.

DICE project in \cite{Guerriero2016}\cite{Artac2018} presents a DSM offering big data design which comprises data, computation, technology-frameworks, and deployment concepts to design and deploy data- intensive applications. DICE proposes a model-driven engineering approach to develop application models which are automatically transformed into IaC. In addition, DICE includes quality of service requirements associated to elements within the application, which are analogous to QS. 
Perez et al. in \cite{perez2019} presented a profile to enable performance and reliability assessment. DICE supports configuration management, service provisioning, and application deployment, but technology-neutral models and architectural tactics are not considered which could hinder portability and design decision tracing. Due to its focus, DICE requires design at very detailed level, specifying different constructs regarding target technologies, but in our proposal, the technology-specific generators transform functional and deployment artifacts to code.

To summarize, the related work approaches reviewed tackle the BDA applications design, but they are not concern about deployment architectural decisions. Specifically, only four proposals follow the SoC principle (\cite{Alrifai2014},\cite{OpenDataGroup},\cite{Anandan2015},\cite{Guerriero2016}), and among them, only Qualimaster and DICE \cite{Guerriero2016} offer a deployment viewpoint. From the architecture perspective, tactics and QS specifications are scarcely ever considered. Based on these findings, we argue that our proposal aims to bridge such gaps. 

\section{ACCORDANT: A DevOps and Domain Specific Model Approach}
\label{sec:proposal}

\begin{figure*}[ht]
\centering
\includegraphics[width=1.0\textwidth]{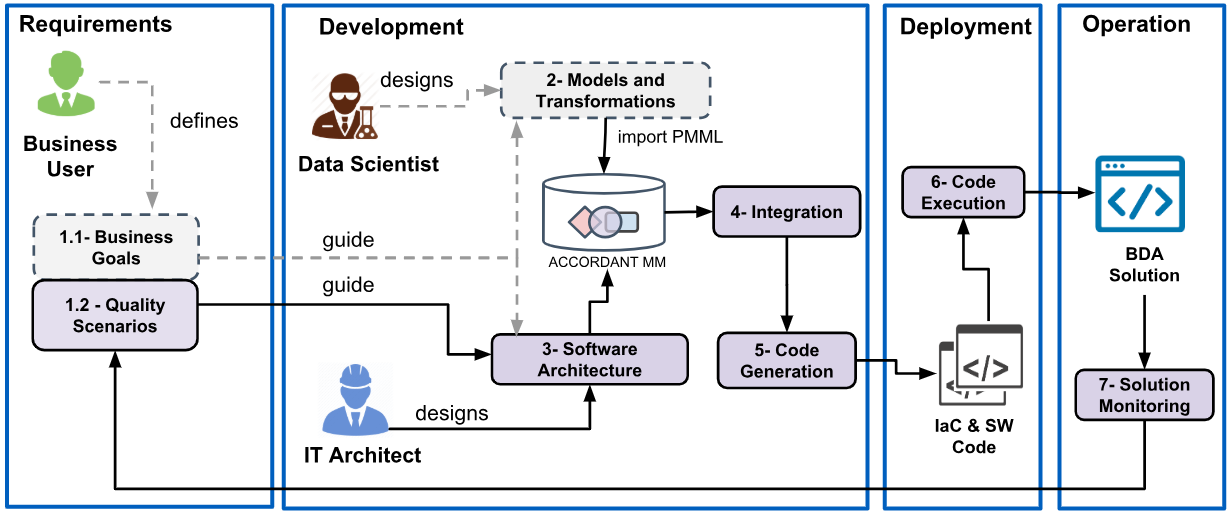}
\caption{ACCORDANT Process overview}
\label{fig:process}
\end{figure*}

This proposal aims at offering a high-level approach to design BDA solutions starting from architectural artifacts, instead of source code. Specifically, we propose ACCORDANT (An exeCutable arChitecture mOdel foR big Data ANalyTics) to deal with functional, infrastructure and QS requirements. Our proposal comprises: a design and deployment process, and a DSM framework to support such process. This paper extends metamodel proposed in \cite{Castellanos2018} by aligning ACCORDANT process to ADD, and including architectural inputs, containerization and serverless deployments in DV. 
Fig.~\ref{fig:process} depicts the ACCORDANT's process, which adapts and integrates an architecture design method (ADD) and analytics methodologies. 

The steps performed using ACCORDANT modeling framework are framed in solid lines, while the steps made with external tools are represented by dotted lines. ACCORDANT process is iterative and, it is composed of seven steps: the business user defines 1.1) business goals and 1.2) QS which will guide the next steps. 2) The data scientist develops data transformations, build and evaluates analytics models. The resulting analytics models are exported as PMML files. 3) Architect design the software architecture using ACCORDANT Metamodel in terms of \textit{Functional Viewpoint}(FV) and \textit{Deployment Viewpoint}(DV). FV model makes use of PMML models to specify the software behavior. 4) FV and DV models are interweaved to obtain an integrated model. 5) Code generation of software and infrastructure is performed from integrated models. 6) The code generated in the previous step is executed to provision infrastructure and install the software. 7) QS are monitored in operation to be validated, and design adjustments can be made to achieve QS, if necessary.

\subsection{Architectural Inputs}

\begin{figure*}[ht]
\centering
\includegraphics[width=1.0\textwidth]{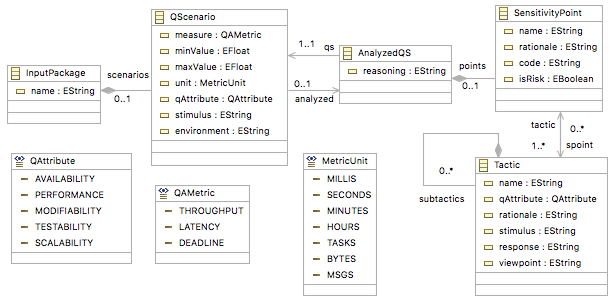}
\caption{Excerpt of Architecture Inputs metamodel.}
\label{fig:rq}
\end{figure*}

According to architecture design methods such as Attribute-Driven Design (ADD)\cite{wojcik2006}, architecture design is driven by predefined quality scenarios (QS) which must be achieved through design decisions compiled in well-known catalogs of architectural patterns and tactics. Both QS and tactics are inputs of the architecture design, therefore we include these initial building blocks in the ACCORDANT metamodel along with other concepts defined in ADD. Fig.~\ref{fig:rq} details the main input building blocks grouped by the architectural input package (\textit{InputPackage}) which contains the elements required to start the architectural design: Quality Scenario (\textit{QScenario}), Analyzed QS (\textit{AnalyzedQS}), \textit{SentivityPoint} and \textit{Tactic}. 
A \textit{QScenario} determines a quality attribute requirement (i.e. latency, availability, scalability, etc) for a specific \textit{Artifact}. Thus, for instance, a QScenario could be defined as ``latency $<=$ 3 seconds for an artifact \textit{X}'', where artifact \textit{X} corresponds to a software component or connector. A QS is analyzed through a \textit{AnalyzedQS}, and sensitivity points. A \textit{SensitivityPoint} is a property of a decision (a set of elements and their relationships within an architectural view) that is critical for achieving the QS, and that such decision is the application of a tactic to a specific application context. Finally,\textit{Tactic} elements synthesize BDA tactics found in \cite{Gorton2014,ullah2019} to be applied in an architecture instance, e.g.: dynamic resource allocation, health monitoring, parallel processing, feature selection, etc.

Once \textit{QScenarios}, \textit{AnalyzedQS}, and \textit{SensitivityPoints} are defined in the \textit{step 1.2} of ACCORDANT process, the software architecture is designed in \textit{step 3} and expressed on the views instantiating tactics in a concrete application. These decisions are associated via \textit{SensitivityPoints}, and they will be evaluated against the initial QS to validated whether the architecture is achieving its goal.

\subsection{Functional Viewpoint (FV)}

\begin{figure*}[ht]
\centering
\includegraphics[width=1.0\textwidth]{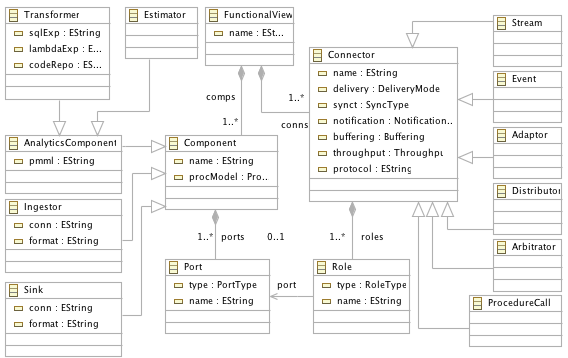}
\caption{Excerpt of Functional Viewpoint of ACCORDANT metamodel.}
\label{fig:fv}
\end{figure*}

FV allows us to design analytics pipelines in terms of ingestion, preparation, analysis and exporting building blocks. \textit{FV} specifies functional requirements of the analytics solution, and the constructs are described in a technology-neutral way as detailed in the metamodel depicted in Fig.~\ref{fig:fv}. FV is expressed in a component-connector structure. Sensitivity points, from architectural inputs, can be associated to components and connectors to represent where architectural decisions have impact regarding the QS. Component metaclasses are specialized in \textit{Ingestors}, \textit{Transformers}, \textit{Estimators} and \textit{Sinks}. \textit{Estimators} and \textit{Transformers} are the software component realizations of PMML data model and data transformer respectively, and the PMML file defines their behavior. A \textit{Component} exposes required and provided \textit{Port}.

\textit{Connectors} metaclasses transfer data or control flow among components through an input or output \textit{Roles}. A set of connector types are defined based on the connector's classification proposed by Taylor et al. in \cite{taylor2010}: \textit{Procedure Call}, \textit{Event}, \textit{Stream}, \textit{Adaptor}, \textit{Distributor} and \textit{Arbitrator}. A \textit{Procedure Call} connector models the flow control and communication through invocations. Similarly, an \textit{Event} connector affects the control flow and provides data transfer, but it is subject to the occurrence of events to notify all interested parts. A \textit{Stream} connector is used to perform transfer of large amounts of data that is continuously generated. \textit{Adaptors} enable interaction between components that have not designed to interoperate providing conversion features. \textit{Distributor} connectors identify interaction paths and communication routing. An \textit{Arbitrator} streamlines system operation and resolves conflicts thus offering intermediary services.

\subsection{Deployment Viewpoint (DV)}

\begin{figure*}
\centering
\includegraphics[width=1.0\textwidth]{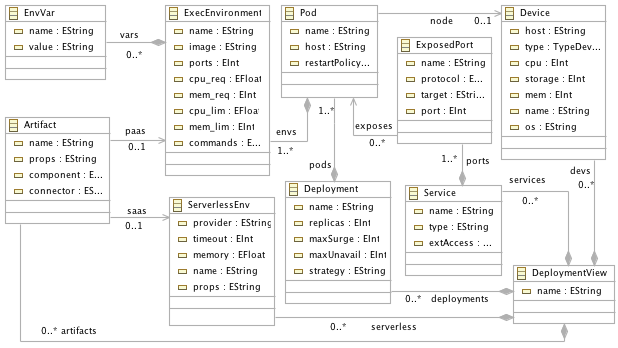}
\caption{Excerpt of Deployment Viewpoint metamodel}
\label{fig:dv}
\end{figure*}

The Deployment viewpoint integrates DevOps practices including containerization, IaC, and serverless computing. The DV specifies how software artifacts (components and connectors) are deployed on a set of computation nodes. The main metaclasses are detailed in Fig.~\ref{fig:dv}. DV metamodel comprises \textit{Pod}, \textit{ExposedPort}, and \textit{Deployment} metaclasses to operationalize BDA applications in a specific technology. It is noteworthy that a \textit{FV} model can be deployed in different \textit{DV} models either to use a different strategy, or to test the fulfillment of predefined QScenarios.

DV contains \textit{Devices}, \textit{Services}, \textit{Deployments}, serverless environments (\textit{ServerlessEnv}), and \textit{Artifacts}. Sensitivity points can be assigned to Deployments and Artifacts to map critical architectural decisions in the DV. A \textit{Device} is a worker machine (physical or virtual) on which the Pods are deployed. A \textit{Pod} is a group of one or more execution environments (\textit{ExecEnvironment}) which can share storage and network. An \textit{ExecEnvironment} represents a container with a Docker image, and specific resources requirements (CPU, memory). A \textit{Deployment} specifies the desired state for a Pod's group and its deployment strategy, including the number of replicas. \textit{Services} and \textit{ExposedPorts} define the policies, addresses, ports, and protocols by which to access to Pods from outside the cluster network. A \textit{ServerlessEnv} element describes a computing environment in which a cloud provider runs the server, and dynamically manages the allocation of machine resources, as opposition to \textit{ExecEnvironment} where physical resources have to be defined and managed. \textit{Artifacts} correspond to executable or deployable representations of functional elements (i.e. components and connectors from functional view) which can be deployed on either execution or serverless environments.

Once PMML, FV and DV models are designed and integrated, code generation takes place by means of model-to-text transformations. Code generation is twofold: software and infrastructure (IaC) code. On the software side, each component and connector is assigned to a specific technology regarding its constraints specified in the model (processing model, ML algorithm, delivery type, sync type, etc). Such assignment enables us to generate code for target technology restricted to these constraints. For instance, near real-time analytics requires stream or micro-batch processing offered by Apache Storm or Spark respectively, and Event connectors such as Apache Kafka or RabbitMQ. Regarding the QS monitoring, code generators include specific machinery to log metrics at an application level. It allows us to collect specific-QS from a high-level abstraction, saving the cost of adding code for logging metrics for each application and target technology. 
On the IaC side, DV model is transformed into Kubernetes' configuration files (in YAML format) used to create and configure infrastructure over Kubernetes cluster. Kubernetes files contain Nodes, Pods, Deployments, and Services which are executed through Kubectl. 

In the last step, the performance metrics of the BDA application are gathered to be compared to initial QS and evaluate the fulfillment of quality requirements. In this step, the architect has to check the outputs, and to make decisions in the architectural views, if QS is not achieved.  This process can take several iterations, and this is the whole cycle that we expect to accelerate and using ACCORDANT.

\section{Evaluation with four BDA Use Cases}
\label{sec:evaluation}

Our experimentation aims to compare development and deployment time for each iteration using accordant and other two frameworks reviewed in Section \ref{sec:relatedwork}: FastScore and SpringXD. We chose these frameworks because they are the closest to our approach, and they support portable analytics models (PMML or PFA). We validated our proposal in different domains through four use cases: UC1) Transport delay prediction, UC2) Near mid-air collision detection, UC3) Near mid-air collision risk analysis, and UC4) El Nino/Southern Oscillation cycles. Table~\ref{tab:usecases} summarizes the use cases, domains, processing models and quality attributes. 
These use cases are applied to analytics models, they also illustrate BDA facets as streaming and micro-batch to deal with the velocity aspect, and batch processing is focused on volume, in terms of data size and computation complexity. 
Fig.~\ref{fig:swarch} details the component-connector model for each use case to illustrate the functional building blocks, and their composition as BDA pipelines. The ACCORDANT specification of these use cases is publicly available\footnote{http://github.com/kmilo-castellanos/accordant-usecases}, and the use cases description will be presented below.

\begin{table}[]
\caption{Use Cases}
\label{tab:usecases}
\begin{adjustbox}{max width=\textwidth}
\begin{tabular}{l>{\raggedright\arraybackslash}p{3cm}lp{2.cm}p{2cm}p{2cm}}
\toprule
Use Case & Description & Domain & Analytics Model & Processing Model & QS Metric \\ 
\midrule
UC1 & Transport Delay Prediction & Transportation & Regression Tree & Stream & Update time, Latency \\
UC2 & NMAC Risk \mbox{Analysis} & Avionics & K-means & Batch & Deadline\\
UC3 & NMAC Detection & Avionics & Decision Tree & Micro-batch & Latency\\
UC4 & El Nino/Southern Oscillation & Weather & Polynomial Regression & Batch & Deadline \\ 
\hline
\end{tabular}
\end{adjustbox}
\end{table}

\subsection{Use Case 1 (UC1)}

The first use case was presented in \cite{Castellanos2018}, and it deals with delay prediction of public transportation in Vancouver. Bus trips data is collected in real-time from Vancouver Transport Operator, and it contains bus stops, routes and time. A regression tree model to predict bus delays (in seconds) is built, evaluated, and exported to PMML. The pipeline, described in Fig.~\ref{fig:swarch}a, starts with an ingestor component which receives HTTP request and put it into an event connector (message broker), then the request message is consumed by the estimator to predict the delay time, and queue it, to be stored into a No-SQL database (hierarchical).
The PMML model is deployed into productive environment as a delay predictor service, using OpenScoring, and Kafka message broker, and MongoDB writer as target technologies. 
The QS were defined in terms of performance and modifiability attributes. The QS specifies that users make 1000 requests to delay prediction service under operations without load, and the responses must have an average latency lower than 2 seconds.
Second QS states that when data scientist produces a new version of the predictive model (new PMML file), it must be updated at runtime within 10 seconds.

\subsection{Use Case 2 (UC2)}
 
UC2 was applied in aviation safety to detect near mid-air collisions (NMAC) on different air space ranges with different deployment models while performance QS are monitored. This use case is described in Fig.~\ref{fig:swarch}b), and it was presented in \cite{Castellanos2019}. NMAC detection comprises a pairwise comparison of flights: $C^2_{n}$), where \textit{n} is the number of flights. Each comparison requires to calculate distance and time based on location, speed and heading to determine the risk level of NMAC, which implies an intensive computation of quadratic time complexity. Eight-hours of data were stored in a distributed file system to be loaded by JSON reader component. This ingestor calls NMAC detector which computes the alert level. Once an alerting level is calculated for each flight pair, the results are sent to the clustering estimator to be associated with a specific cluster. NMACs are stored back in the file system. To compare different data size magnitudes, we collected flight data for three air space ranges in nautical miles (nmi): 2 nmi, 20 nmi, 200 nmi, and 1500 nmi around John F. Kennedy Airport. These ranges represent different application scopes to attend various demand levels: local, metropolitan, and regional areas. The largest dataset (1500 nmi) is 1.4 GB of JSON files. This use case did not have real-time requirements due to its heavy workload nature, and therefore a performance QS for deadlines lower than one hour was defined.

\subsection{Use Case 3 (UC3)}

\begin{figure}
\centering
\includegraphics[width=0.8\columnwidth]{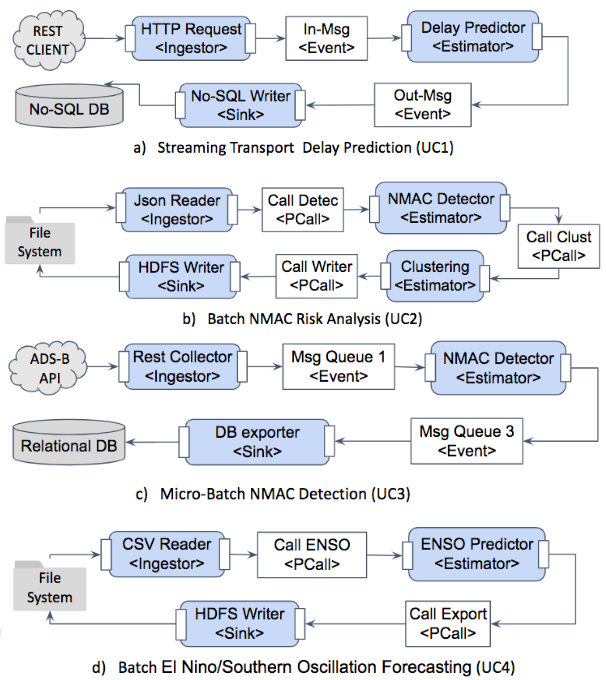}
\caption{Component diagrams of Use Cases}
\label{fig:swarch}
\end{figure}

UC3 is a real-time application to detect NMAC within an air space range, and its architecture is described in Fig.~\ref{fig:swarch}c).  The ingestor component consumed data through direct REST service. Flight data was pushed in a message queue to be consumed by the NMAC detector component which performed the potential collision detection to be finally stored in a relational DB through a message broker connector. It is worth mentioning that the NMAC estimator of UC2 and UC3 are the same, since its inputs, outputs, and behavior are identical, so we can reuse such functional component definition, in spite of its deployment can be different regarding the QS constraints.
Given the near real-time nature of this application, latency is the critical quality attribute, and we evaluated this metric in two ranges of air space around John F. Kennedy Airport: 2 nmi and 200 nmi, which demand different computation resources.

\subsection{Use Case 4 (UC4)}

In this last use case, we used a public available data and PMML model (polynomial regression) of El Nino/Southern Oscillation (ENSO)\footnote{http://dmg.org/pmml/pmml\_examples} to implement a batch oriented pipeline, see Fig.~\ref{fig:swarch}d. The El Nino/Southern Oscillation (ENSO) cycle, was the strongest of the century which produced many problems throughout the world affecting South and North America countries with destructive flooding in some areas and strong drought in other areas. Data for this use case contains oceanographic and surface meteorological readings (geolocation, humidity, surface winds, sea surface temperatures, and subsurface temperatures) are taken from a series of buoys positioned throughout the equatorial Pacific. This data is expected to help with the understanding and prediction of ENSO cycles. We read the historic data from 1980 to 1998 (178,080 records) using a CSV reader (ingestor) component, which sends the data to the ENSO predictor component. ENSO predictor is a estimator component that forecasts air temperature, and stores the prediction in a distributed file system. The QS defined for UC4 was a deadline for batch processing lower than 30 minutes.

\subsection{Development, Deployment Time and Gain Factor}
\label{sub:gf}
To compare ACCORDANT, SpringXD and FastScore, we measured the time invested in development and deployment phases for each use case. \textit{Development phase} involves design and development of the functional components and connectors in a specific technology. \textit{Deployment phase} comprises the design and provisioning of the technology infrastructure, the installation of software artifacts developed in the previous phase, and the monitoring of the solution regarding the predefined QS. These phases are performed iteratively, since in each iteration some improvements and refinements are done until the QS are achieved. Therefore, we measure the time invested in each iteration, and also we calculate the gain factor $GF(uc,f)$, as a metric to estimate the cumulative average of time reduction ratio for a use case \textit{uc}, using framework \textit{f} over \textit{I} iterations. $GF(uc,f)$ is defined as follows:

\begin{equation}
GF(uc,f) = \dfrac{1}{I} \sum_{i=1}^{I} \dfrac{time\_spent(uc,f)_{i} - time\_spent(uc,f)_{i+1}}{time\_spent(uc,f)_{i}}
\label{eq:dde}
\end{equation}

We define the gain factor as a form to measure the incremental improving of using a high level abstractions to modify or refine an application until achieve an expected QS. The time for each use case, phase, and iteration was collected from two development teams which learnt and used the three frameworks to develop and deploy two use cases each one, while they were recording the time spent. The development and deployment process using ACCORDANT will be illustrated with UC4 in the next Section.

\section{Experimentation}
\label{sec:experimentation}

To design, develop and deploy the four use cases, we followed ACCORDANT process detailed previously in Figure~\ref{fig:process}. For the sake of brevity, this section details the step-by-step implementation of UC4 as an example, more details about the other use cases can be found in \cite{Castellanos2018,Castellanos2019}. The ACCORDANT projects are available in a public repository \footnote{http://github.com/kmilo-castellanos/accordant} as well as use cases and results \footnote{http://github.com/kmilo-castellanos/accordant-usecases}.

\subsection{Definition of Quality Scenarios}
\label{subsec:qs}

\begin{figure*}
\centering
\includegraphics[width=1.0\textwidth]{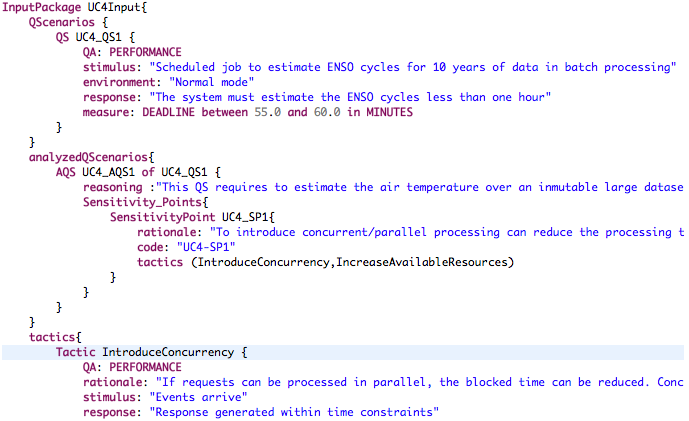}
\caption{Excerpt of Input Package Models of UC4 Using ACCORDANT DSLs}
\label{fig:rq_uc4}
\end{figure*}

QS are defined regarding the use case's quality requirements. In UC4, a scheduled job to estimate ENSO cycles for ten years of data is processed in batch. In this vein, Fig~\ref{fig:rq_uc4} details architectural inputs of UC4 expressed using the ACCORDANT's input package DSL. The predictor component is required to have a deadline lower than 1 hour in the QS \textit{UC4\_QS1}. Analyzing this QS, a sensitivity point (\textit{UC4\_SP1}) is identified to achieve the deadline metric by applying two tactics: \textit{introduce concurrency} and \textit{increase}. available resources. These tactics will be materialized in the software architecture design. 

\subsection{Development of Data Transformations and Analytics Model}
\label{subsec:analyticsmodel}

\begin{figure*}
\centering
\includegraphics[width=0.8\textwidth]{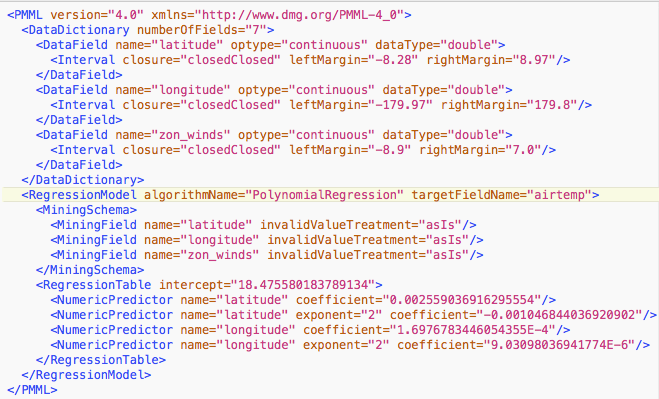}
\caption{Excerpt of ENSO Polynomial Regression Model of UC4 in PMML format}
\label{fig:ensopmml}
\end{figure*}

Analytics Model is trained and evaluated by the data scientist outside the ACCORDANT framework, and the resulting models were exported to PMML file to be loaded in the ACCORDANT functional model. In this case, the polynomial regression model of ENSO is downloaded and used. Fig~\ref{fig:ensopmml} describes the structure of the PMML, detailing some data fields, mining fields, and regression coefficients. This PMML file will be embedded in the functional model in the next step.

\subsection{Design of Software Architecture - Functional View}
\label{sec:sw_design}

\begin{figure}
\centering
\includegraphics[width=1.0\columnwidth]{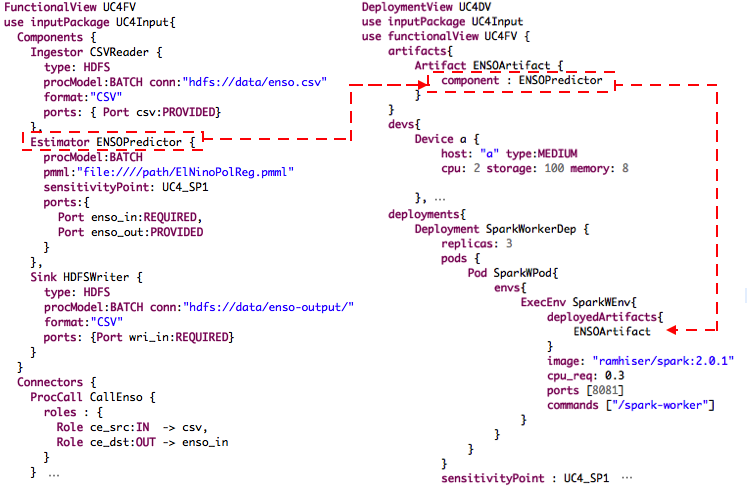}
\caption{Excerpt of Functional (a) and Deployment (b) Models of UC4 Using ACCORDANT DSLs}
\label{fig:fv-dv_model}
\end{figure}

FV models were designed using ACCORDANT Functional DSL to specify a component-connector structure for each use case. Two iterations of functional model were designed for UC4, and the last iteration is depicted in Figure~\ref{fig:fv-dv_model}a. Since architectural inputs are required in this design, this package is imported using the keyword \textit{use inputPackage}. The functional model specifies three components: (\textit{CSVReader::Ingestor}, \textit{ENSOPredictor::Estimator}, and \textit{HDFSWriter}::Sink), and two connectors: procedure calls \textit{CallEnso::ProcCall} and \textit{CallExport::ProcCall} which connect the components through ports. The components also include some properties such as connections and formats. Additionally, \textit{ENSOPredictor} uses batch processing model, it has associated the PMML ``ElNinoPolReg.pmml'', obtained in the previous step, to provide the predictive behavior. The sensitivity point \textit{UC4\_SP1} aligns the architectural input (QS and tactics explained in Section \ref{subsec:qs}) to ENSOPredictor. It means that \textit{ENSOPredictor} becomes part of the \textit{introduce concurrency} tactic realization that will be translated into a distributed processing model which has to be supported by the target technology.

\subsection{Design of Software Architecture - Deployment View}
\label{subsec:dvexp}

The deployment view models were designed using ACCORDANT DSL for each use case defined in the functional models. The UC4 deployment model had three iterations, and Fig.~\ref{fig:fv-dv_model}b details the last version. Given that DV is based on input package and functional view, they are imported by means of keyword \textit{use inputPackage} and \textit{functionalView} respectively. 
This view includes the artifacts that map connectors and components from functional view (e.g. \textit{ENSOPredictor}) to deployable elements (e.g. \textit{ENSOArtifact}). Devices and deployments were specified to support the computation requirements. For instance, deployments of Spark master and worker nodes (e.g. \textit{SparkWorkerDep}) details the number of replicates, pods and execution environments (\textit{ExecEnv}). ExecEnv defines the docker image, CPU and memory requirements, ports, and commands along with the artifacts to be deployed (\textit{ENSOArtifact}). Finally, the sensitivity point \textit{UC4\_SP1} associates the deployment SparkWorkerDep to performance QS, and the tactic \textit{increase available resources} (see Section \ref{subsec:qs}) to support distributed computing over a Spark cluster.

\subsection{Integration and Code Generation}

Once the FV and DV models were designed and integrated, the code generation produced both the functional code and IaC. On the one hand, the functional code is a Spark driver program as detailed in Listing~\ref{lst:sparkcode}, where \textit{ENSOPredictor} component implements the PMML model in Spark technology. The Spark program defines data input and output from the Data Dictionary and Mining Schema embedded in PMML specifications. On the other hand, infrastructure code is the configuration files which specify the provision and configuration policies of Kubernetes cluster. Listing~\ref{lst:yamlsfile} shows an example of generated Kubernetes files. The whole code of use cases is publicly available in the accordant-usecases repository.

\begin{lstlisting}[basicstyle=\tiny, caption={Generated Java Code of EnsoEstimator Component for Spark Streaming},label={lst:sparkcode}, language=java]

SparkSession sparkSession = new SparkSession(sc.sc());
InputStream pmmlFile = new URL("file:////path/ElNinoPolReg.pmml")
EvaluatorBuilder b = new LoadingModelEvaluatorBuilder().load(pmmlFile);
Evaluator eval = builder.build();
TransformerBuilder pmmlTransformerBuilder =
new TransformerBuilder(evaluator)
    .withTargetCols().exploded(true);
List<StructField> fields = new ArrayList<StructField>();
fields.add(DataTypes.createStructField("latitude", DataTypes.DoubleType, true));
...
fields.add(DataTypes.createStructField("s_s_temp",DataTypes.DoubleType, true));
StructType schema = DataTypes.createStructType(fields);
Transformer pmmlTransformer = pmmlTransformerBuilder.build();
Dataset<Row> inputDs = sparkSession.read().schema(schema).csv("data/Elnino.csv");
TransformerBuilder tb = new TransformerBuilder(eval);
Transformer transformer = tb.build();
Dataset<Row> resultDs = transformer.transform(inputDs);
resultDs.write().option("header", "true").csv("/enso-output/");
...
\end{lstlisting}

\begin{lstlisting}[basicstyle=\tiny,caption={Generated YAML Code from Deployment Specification for Kubernetes (Extract)},label={lst:yamlsfile}]

apiVersion: apps/v1
kind: Deployment
metadata:
  name: SparkWorkerDep
spec:
  replicas: 3
    spec:
      containers:
      - name: SparkWEnv
        image: ramhiser/spark:2.0.1
        command: [/spark-worker]
        ports:
        - containerPort: 8081
        resources:
          requests:
            cpu: 0.3
...        
            
\end{lstlisting}

\subsection{Code Execution}

Kubernetes code was executed on the AWS cloud using Amazon Elastic Container Service for Kubernetes (Amazon EKS) and Elastic Compute Cloud (EC2). After that, the software code was installed over the cluster to operationalize the end-to-end solution.

\subsection{Solution Monitoring}

Performance metrics for each use case in operation were collected and validated against QS defined in Section~\ref{subsec:qs}. As a result, different deployment configurations were designed, deployed and monitored in each iteration to observe the fulfillment of QS.

\section{Results and Discussion}
\label{sec:results}

Revisiting the related work reviewed in Section \ref{sec:relatedwork}, we have shown in practice how ACCORDANT bridge the gap among analytics, software architecture, and DevOps. As presented in Fig.~\ref{tab:rw}, ACCORDANT follows the SoC principle by means of three different languages to specify domain concerns. Analytics models in ACCORDANT are cross-industry and technology-neutral. In terms of software architecture, ACCORDANT supports QS specifications aligned to FV and DV, and these models can be specified independently but in an integrated way. Architectural tactics enable software architects to describe and communicate their decisions. Code generators offer flexibility and impact positively the development and deployment efficiency. Respecting DevOps practice, deployment models allow us to design deployment diagrams, not limited to a number of instances. Continuous deployment is supported via IaC and code generation, and QS-monitoring is implemented by injecting logging code in the generated applications. Finally, self-adaptation is not covered in the current version of ACCORDANT.

To summarize, though a large variety of component-connector metamodels have been previously proposed, as far as we know, our contribution resides in specialize a component-connector metamodel in the BDA domain, and integrate it with architectural inputs and deployment models to offer a holistic design. Additionally, this section presents and discusses the experimental results obtained during the iterative development and deployment phases of UC1, UC2, UC3, and UC4.

\subsection{Development and Deployment Time}

\begin{figure}
\centering
\includegraphics[width=1.0\columnwidth]{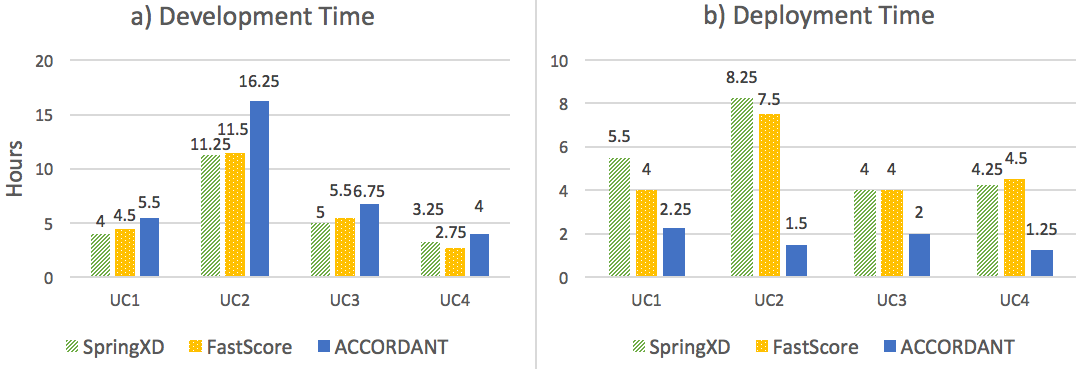}
\caption{Development and Deployment Time for Use Case}
\label{fig:uctimes}
\end{figure}

Fig.~\ref{fig:uctimes} depicts the development and deployment time (in hours) accumulated for all iterations per use case. It is worth noting that development time using ACCORDANT is higher (between 23\% and 47\%) compared to SpringXD and Fastscore, but the deployment time is significantly lower (between 50\% and 81\%) using ACCORDANT. The higher development time can be explained by the time required in ACCORDANT to specify architectural inputs, and many details in the FV. In addition, the current version of the ACCORDANT prototype generates functional code for estimators, but ingestor, sinks, and connectors still require manual. Although ACCORDANT required more effort in the development phase, this effort is rewarded during the deployment phase, where infrastructure and QS-monitoring are provided automatically aligned to Inputs and FV, unlike other approaches. This benefit can be observed on the deployment time across all use cases using ACCORDANT, because they are more similar than the other approaches.

The biggest time differences arise from UC2 that demands more time because it includes a more complex pipeline involving two estimators: NMAC detector and K-means clustering.
Another interesting finding was that the high-level reuse of previous architectural decisions (tactics) reduced the time of development as shown the marked decreasing between use cases, and the growing gain factor among iterations detailed in Fig.~\ref{fig:uctimes}. These results suggest that ACCORDANT is most suitable for application involving multiple iterations, or in subsequent applications where reusing architectural decisions, models, and metrics can reduce development times.

\subsection{Gain Factor Comparison}

\begin{figure}
\centering
\includegraphics[width=1.0\columnwidth]{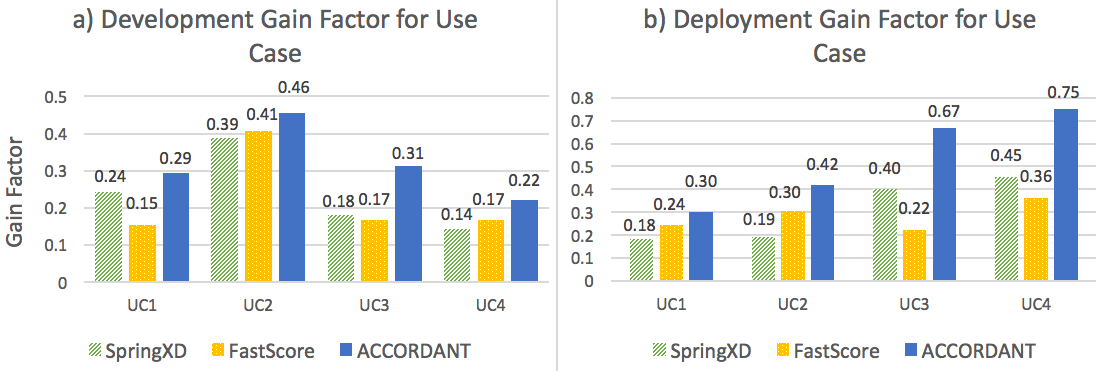}
\caption{Gain Factor for Use Case}
\label{fig:gf}
\end{figure}

The gain factor metric presented in Section \ref{sub:gf} was calculated for each use case and iteration of development and deployment phases as depicted in Fig.~\ref{fig:gf}. ACCORDANT's gain factor was higher for all use cases, in the development phase (Fig.~\ref{fig:gf}a), what suggests that the high-level abstractions promote the highest reduction of development time among consecutive iterations. The highest gain factor was 0.46 in the UC3, it means reducing in 46\% the development time between consecutive iterations. The greatest gain factor difference over the other approaches was 0.13 in the UC3.
Regarding the deployment gain factor (Fig.~\ref{fig:gf}b), ACCORDANT also exhibited the highest gain factor, on an even higher proportion, up to 0.75 in UC4. This means each deployment iteration reduces the time in 75\% compared to the previous one. Similar to the deployment time in the previous section, we argue that the gain factor in the deployment phase is greater because of the IaC generation is not present in the other approaches.

\section{Conclusions}
\label{sec:conclusions}
We have presented a DevOps and DSM proposal to design, deploy, and monitor BDA solutions. We have positioned the ACCORDANT contributions within the related work.
Four use cases from different domains were used to evaluate our approach against two BDA frameworks. As a result, ACCORDANT has shown to facilitate and accelerate iterative development and deployment phases by offering an integrated and high-level design BDA applications. The greatest time reduction was reported in the deployment phase, achieving up to 81\% compared to other approaches. In contrast, the development times offered by ACCORDANT were greater. Despite the longer development time, deployment time is significantly reduced thanks to the QS, FV, and DV alignment. ACCORDANT's gain factor was higher, which implies a higher reduction time in each iteration. 

In contrast, some limitations have emerged from experimentation. The development phase is slower than the other approaches for multiple reasons. The current version of the ACCORDANT's prototype requires supplementary manual coding what increases the development time. ACCORDANT also requires more design details and architectural inputs. These additional definitions are rewarded in consecutive iterations, so ACCORDANT is most suitable for application involving multiple iterations. Finally, our approach takes advantage of reusing architectural decisions and models, hence, first-time or one-time applications may not be benefited from our proposal.

As future work, the performance metrics collected along with FV and DV models could allow us to propose a performance model to predict the expected application-specific behavior based on the functional model, deployment model, and target technology to recommend optimal architecture configuration for a defined QS.
Furthermore, we could include features to simulate and verify correctness properties over the models such as technology selection in the FV model and resource allocation in the DV model. Given that PMML provides a model verification schema to validate results accuracy, a future extension could incorporate automated model verification. This approach has been used for deploying analytics components and connectors on virtual machines over cloud infrastructure, but different paradigms such as serverless or fog computing may open new research lines. 

\pagebreak
\appendix
\section{Abbreviations}
\begin{table}[h]
\caption{Abbreviations}
\label{tab:abbreviations}
\begin{tabular}{l|l}
\hline
ADD & Attribute Driven Design Method \\
ADS-B & Automatic Dependent Surveillance—Broadcast \\
BDA & Big Data Analytics \\
DSL & Domain Specific Language \\
DSM & Domain Specific Model \\
DV & Deployment View \\
ENSO & El Nino/Southern Oscillation \\
FV & Functional View \\
IaC & Infrastructure as Code \\
NMAC & Near Mid-air Collision \\
PFA & Portable Format for Analytics \\
PMML & Predictive Model Markup Language \\
QS & Quality Scenario \\
\hline
\end{tabular}
\end{table}

\bibliography{mybibfile}

\end{document}